# Mixed-dimensional topological charge compensation beyond Weyl-node pairing

Xiao-Jing Gao, Ke-Xin Pang, Yun-Yun Bai, Yanfeng Ge, and Yan Gao*

*State Key Laboratory of Metastable Materials Science and Technology & Hebei Key Laboratory of Microstructural Material Physics, School of Science, Yanshan University, Qinhuangdao 066004, China*

**The Nielsen–Ninomiya no-go theorem requires the total topological charge ($C$) in a periodic Brillouin zone to vanish, conventionally leading to opposite-chirality pairs of Weyl points (WPs). A charged nodal surface (NS) provides an alternative mechanism for compensating a same-chirality Weyl pair, yet the general symmetry constraints governing this configuration and its electronic realization remain unexplored. Here, by an exhaustive analysis of all 1651 magnetic space groups (MSGs), with and without spin–orbit coupling, we identify only 13 type-II MSGs that permit the realization of a minimal unpaired mixed-dimensional configuration comprising a single $|C| = 2$ NS and two same-chirality $|C| = 1$ WPs. Guided by this symmetry framework, we identify the chiral boron allotrope THC-$B_{30}$ as the first electronic realization of this unconventional topology. First-principles calculations reveal only two WPs with $C = -1$ and one NS with $C = +2$, satisfying global topological charge neutrality and exhibiting unconventional Fermi arcs connecting the WP and NS projections. Moreover, strain-induced topological phase transitions redistribute topological charge among distinct band-degeneracy manifolds. Our findings establish a symmetry-guided route to realizing same-chirality Weyl pairs through mixed-dimensional charge compensation beyond conventional Weyl-node pairing.**

*Corresponding author: yangao9419@ysu.edu.cn

Weyl points (WPs) are monopoles of Berry curvature whose quantized chiral charges govern the bulk topology and Fermi-arc surface states of Weyl semimetals (WSMs)[1-4]. The Nielsen–Ninomiya no-go theorem requires the total chiral charge in a periodic Brillouin zone (BZ) to vanish[5, 6]. When all topological charges are carried by isolated WPs, the minimal charge-neutral configuration consists of two WPs with opposite chiralities[7-12]. However, time-reversal symmetry relates WPs at opposite momenta while preserving their chirality[1]. Consequently, a single pair of same-chirality unit-charge WPs carries a net topological charge of ±2, which cannot be compensated within the pair itself. Whether such a Weyl pair can constitute the only point nodes of a charge-neutral topological semimetal therefore depends on the existence of an alternative charge-compensation mechanism.

Topologically charged nodal surfaces (NSs) provide such an alternative by extending Berry-curvature monopoles from isolated points to two-dimensional (2D) band-degeneracy manifolds[13-17]. In particular, nonsymmorphic screw rotations combined with time-reversal symmetry can enforce NSs at BZ boundaries[18], which may carry nonzero Chern numbers ($C$) and compensate the net charges of isolated WPs[15-17]. This mechanism has been established in artificial acoustic crystals, where a charge-two NS coexists with two same-chirality unit-charge WPs and gives rise to unconventional surface-state arcs connecting their projections[15-17]. In electronic systems, charged nodal walls have been experimentally observed in PtGa[19], where they compensate the net charge of multiple WPs, while AlPt has been shown to host a nodal surface with a giant Chern number[20]. These advances demonstrate that extended nodal manifolds can serve as independent topological-charge carriers, enabling unconventional charge-neutral configurations beyond conventional Weyl-node pairing.

Despite these developments, the general symmetry conditions governing a minimal unpaired composite of one charged NS and two same-chirality WPs remain incompletely understood, particularly in electronic systems. Previous symmetry analysis identified six space groups supporting this configuration in nonmagnetic spinless systems and demonstrated corresponding realizations in phonon spectra[15]. However, a comprehensive classification incorporating both spin–orbit coupling (SOC) and magnetic space-group symmetries remains lacking. Moreover, realizing this minimal unpaired configuration in an electronic material requires not only symmetry-compatible WPs and an NS, but also the absence of additional crossings between the relevant bands throughout the BZ. These constraints pose a substantial challenge to identifying electronic materials that host an isolated mixed-dimensional topological composite. Thus, the complete symmetry conditions and a corresponding minimal unpaired electronic realization remain to be established.

Here, we systematically analyse all 1651 magnetic space groups (MSGs), considering both single-valued representations without SOC and double-valued representations with SOC, and identify only 13 type-II MSGs compatible with a single charged NS and two same-chirality unit-charge WPs. Guided by this symmetry classification and high-throughput structure searches, we predict a chiral boron allotrope, THC-$B_{30}$, with space group $P6_3$ (No. 173), as the first candidate electronic realization of this topological configuration. First-principles calculations reveal that the relevant bands host only two WPs with $C = -1$ along the Γ–A direction and one NS with $C = +2$ on the $k_z = \pi$ plane, forming a minimal unpaired mixed-dimensional three-terminal topological composite that satisfies global charge neutrality. The corresponding (100) surface exhibits unconventional Fermi arcs connecting the WP projections to the NS projection. Furthermore, hydrostatic pressure and uniaxial strain induce topological phase transitions that redistribute chiral charges between WPs and the NS, even driving the NS charge to zero while preserving the NS itself. Our findings establish a symmetry-guided route towards realizing and controlling same-chirality Weyl pairs through nodal-surface topological charge compensation in electronic crystals.

## Results

**Symmetry criteria for a single charge-two nodal surface and a single pair of unit-charge Weyl points.** A minimal unpaired mixed-dimensional configuration comprising one charged NS and two same-chirality unit-charge WPs is subject to both symmetry compatibility and minimal-orbit constraints. Denoting the two WPs by $\boldsymbol{k}_{\mathrm{W1}}$ and $\boldsymbol{k}_{\mathrm{W2}}$, with $C_{\mathrm{W1}} = C_{\mathrm{W2}} = \pm 1$, we require the host MSG $G$ to satisfy the following two conditions.

**(i) Symmetry compatibility.** The MSG ($G$) must allow a symmetry-protected NS and unit-charge WPs to coexist between the same adjacent bands. In the relevant nonsymmorphic systems, the combined twofold screw rotation and time-reversal symmetry can protect a nodal plane at $k_z = \pi$[18]. Although the existence of the NS is symmetry protected, its Chern number is not independently fixed by this degeneracy. When no additional topological-charge carriers occur between the relevant bands, global neutrality requires $C_{\mathrm{NS}} = -(C_{\mathrm{W1}} + C_{\mathrm{W2}}) = \mp 2$.

**(ii) Minimal symmetry orbits.** The NS must remain a single symmetry-invariant manifold, while the two WPs must form a closed same-chirality two-node configuration under $G$. This can occur either when both WPs occupy individually symmetry-invariant momenta, with $G_{\mathrm{W1}} = G_{\mathrm{W2}} = G$, or when they belong to a two-point star related by a chirality-preserving symmetry. In the latter case, $G = G_{\boldsymbol{k}_1} \cup \mathcal{O} G_{\boldsymbol{k}_1}$ and $G_{\boldsymbol{k}_2} = \mathcal{O} G_{\boldsymbol{k}_1} \mathcal{O}^{-1}$, where $\mathcal{O} = \mathcal{R}\ or\ \mathcal{R}\mathcal{T}$ (where $\mathcal{R}$ and $\mathcal{T}$ are proper rotation and time-reversal operators, respectively) maps one WP onto the other without reversing its Chern charge. These constraints exclude additional symmetry-generated nodes and preserve the minimality of the target configuration.

Based on the encyclopedia of emergent particles[21-23], we apply these criteria to all 1651 MSGs, both without and with SOC, and identify only 13 compatible type-II MSGs: Nos. 4, 17, 20, 76, 78, 91, 95, 169, 170, 173, 178, 179, and 182. The classification includes unit-charge WPs on both high-symmetry points (HSPs) and lines (HSLs), extending the six spinless space groups previously identified for WPs restricted to HSPs[15]. Table 1 summarizes the allowed configurations and associated irreducible representations (IRRs) or corepresentations, while the detailed classification is provided in the Supplementary Table S1. These symmetry conditions provide theoretical guidance for the subsequent search for electronic realizations of this unique topological phase.

**Symmetry-guided discovery and structural characterization of THC-$B_{30}$.** Guided by the symmetry criteria, we performed high-throughput structure searches to identify electronic materials capable of hosting the minimal unpaired mixed-dimensional topological composite. Elemental boron was selected because its versatile multicentre bonding supports diverse crystalline configurations[24], while its weak SOC favours the preservation of symmetry-protected band crossings[25-30]. We restricted the searches to the 13 nonmagnetic space groups identified by our symmetry analysis. As illustrated in Fig. 1, the RG2[31] random-structure search generated 13,465 initial structures, of which 2,199 passed the three-dimensional structural screening. Subsequent first-principles electronic-structure calculations identified 77 candidates with symmetry-compatible semimetallic band crossings. Further assessment of phonon and elastic stability, combined with a magnetic-moment criterion of $\mu < 0.1$ μB, reduced the number to 15 candidates. Finally, a systematic analysis of band degeneracies and their topological charges identified one candidate exhibiting the targeted configuration of a single charged NS and two same-chirality unit-charge WPs.

We designate this predicted boron allotrope as triangular helical-chain boron (THC-$B_{30}$). As shown in Fig. 2, THC-$B_{30}$ adopts a chiral hexagonal structure with space group $P6_3$ (No. 173). Its primitive cell contains 30 boron atoms distributed over five inequivalent 6c Wyckoff positions (Supplementary Table S2), with optimized lattice parameters $a = b = 7.55$ Å and $c = 4.76$ Å, and a theoretical density of 2.29 g/cm$^3$ (Supplementary Table S3). The crystal framework consists of interconnected bilayer triangular $B_6$ units and triangular helical boron chains. Each $B_6$ unit comprises two stacked boron triangles with B–B bond lengths of 1.73 and 1.74 Å, respectively [Fig. 2(b)]. The helical units consist of three interconnected chains formed from triangular boron motifs, which wind around the crystallographic *c* axis and give rise to left- and right-handed enantiomeric structures [Figs. 2(c–e)]. This structural chirality provides a crystallographic basis for investigating the relationship between enantiomeric structures and the chirality of their topological band crossings.

The calculated total energy of THC-$B_{30}$ is −6.38 eV/atom, approximately 0.32 eV/atom higher than that of near-ground-state α-$B_{12}$ (−6.70 eV/atom), indicating its metastability relative to α-$B_{12}$. Its energy is also comparable to those of several previously proposed three-dimensional boron allotropes[32-38] (see Supplementary Table S3). The calculated elastic constants satisfy the Born mechanical stability criteria[39], while the phonon dispersions of both enantiomers exhibit no imaginary frequencies along all high-symmetry paths [Supplementary Figs. S1(a,b)]. Furthermore, *ab initio* molecular dynamics simulations at 800 K show that the structural frameworks remain intact over 8 ps, accompanied by bounded fluctuations in the potential energy [Supplementary Figs. S1(c,d)]. Together, these results support the local mechanical and dynamical stability of THC-$B_{30}$ and its finite-time structural persistence under the simulated thermal conditions, providing a candidate material platform for investigating mixed-dimensional topological charge compensation.

**Electronic structure and mixed-dimensional topological charge compensation.** Having established the structural stability of THC-$B_{30}$, we investigate its electronic band structure and the topological properties of the symmetry-protected band crossings. As shown in Fig. 3(a), the 45th and 46th bands cross near the Fermi level along the Γ–A direction, forming a WP, with its time-reversal-related counterpart located along the opposite direction of the same rotational axis. The band degeneracies along A–L–H belong to a 2D nodal surface (NS) extending across the $k_z = \pi$ plane, with crossing energies ranging from −0.23 to 0.12 eV. A systematic examination of the corresponding band gap throughout the BZ identifies only two isolated WPs and a single NS between these adjacent bands [Fig. 3(f)]. The projected density of states in Fig. 3(b) indicates that the electronic states near these crossings are dominated by the $p$ orbitals of boron atoms in the triangular helical chains, particularly the $p_z$ orbital. This orbital character is further supported by the charge-density distribution at the WP [Fig. 3(c)].

To establish the topological nature of the isolated crossings, we examine their symmetry representations and topological charges. As illustrated in Fig. 3(g), the two crossing bands along Γ–A transform according to the irreducible representations $\Delta_1$ and $\Delta_6$, whose sixfold rotational eigenvalues differ by a phase factor of $e^{\pm i\pi/3}$. According to the rotational symmetry criterion for WPs[40], this eigenvalue relation is compatible with a conventional unit-charge WP. The corresponding bands exhibit linear dispersion in all directions near the crossing points. The evolution of Wannier charge centres (WCCs) confirms that both WPs, denoted $W_1$ and $W_2$, carry identical Chern numbers of $C = -1$ [Fig. 3(d)]. Their equal chirality is consistent with time-reversal symmetry, which relates the two WPs at opposite momenta while preserving their monopole charges. These results establish the presence of a single pair of conventional WPs with the same chirality, together with a NS, between the 45th and 46th bands.

In contrast to the isolated WPs, the 2D NS is enforced by the combined action of the twofold screw rotation $S_{2z}$ and time-reversal symmetry $\mathcal{T}$[18]. In the spinless limit, the corresponding antiunitary symmetry $\Theta = S_{2z}\mathcal{T}$ satisfies $\Theta^2 = -1$ on the BZ boundary $k_z = \pi$, enforcing a twofold degeneracy throughout this plane. The 3D band dispersion in Fig. 3(h) confirms that the two bands remain degenerate on the nodal plane and split away from it, except at the isolated WPs. Although this nonsymmorphic symmetry protects the existence of the NS, its Chern number is determined by the global distribution of topological charges. Since the relevant bands contain only two WPs with $C = -1$ and one NS, global charge neutrality requires the NS to carry a compensating charge of $C_{\mathrm{NS}} = +2$. Consistently, the Berry-curvature distribution in Fig. 3(e) exhibits flux emerging from the NS projection and converging towards the two negatively charged WPs. The NS thus acts as a source of Berry curvature, while the WPs serve as sinks, establishing a charge-neutral mixed-dimensional composite composed of one $C = +2$ NS and two $C = -1$ WPs.

The structural chirality of THC-$B_{30}$ provides an additional connection between real-space handedness and momentum-space topology. Spatial inversion maps the left-handed crystal onto its right-handed enantiomer, thereby reversing the signs of the chiral topological charges of the corresponding WPs. As demonstrated by the WCC evolution in Fig. 3(i), the two WPs in the right-handed structure carry $C = +1$, opposite to those in the left-handed structure. The associated Berry-curvature distribution also reverses its flow direction [Fig. 3(j)], indicating that the WPs become sources of Berry curvature while the NS provides the compensating negative charge. Consequently, the two enantiomers realize opposite charge configurations, $\{+2, -1, -1\}$ and $\{-2, +1, +1\}$, while preserving the same mixed-dimensional topology. These findings demonstrate that THC-$B_{30}$ hosts a charge-neutral composite of topological objects with different dimensionalities, with the signs of their chiral charges linked to the handedness of the crystal structure.

**Chiral surface states and unconventional Fermi arcs.** The coexistence of a charged NS and two same-chirality WPs gives rise to an unconventional manifestation of bulk–boundary correspondence. In conventional WSMs, Fermi arcs connect the surface projections of WPs with opposite chiralities. By contrast, the mixed-dimensional topology of THC-$B_{30}$ allows surface states to connect the projections of negatively charged WPs to that of the positively charged NS. To characterize this distinctive boundary-state connectivity, we calculate the surface spectral function of the semi-infinite (100) surface. As shown in Fig. 4(a), the surface energy spectrum along the high-symmetry path $\bar{\Gamma} - \bar{\mathrm{A}} - \bar{\Gamma}'$ exhibits a dispersive surface-state branch within the projected bulk band structure. The corresponding constant-energy contours at $E_1 = -0.10$ eV, coinciding with the WP energy, reveal open Fermi-arc-like features extending from the projected WPs towards the NS projection [Fig. 4(b)]. Unlike the isolated projections of WPs, the NS at $k_z = \pi$ projects onto an extended boundary line along $\overline{\mathrm{A}} - \overline{\mathrm{H}}$ in the (100) surface BZ. The surface-state branches consequently connect point-like Weyl projections to an extended nodal-manifold projection rather than to WPs of opposite chirality. At $E_2 = -0.20$ eV, the surface spectral contours evolve while retaining characteristic arc-like features associated with the WP–NS topology [Fig. 4(c)]. These results are consistent with the mixed-dimensional topological charge compensation established from the bulk electronic structure.

**Strain-induced topological phase transitions and charge redistribution.** The coexistence of WPs and a charged NS offers a distinctive route to manipulating topological charge in crystalline solids. Unlike isolated WPs, whose charges are constrained by their local band dispersions and crystalline symmetries, a charged NS can compensate the collective monopole charges of multiple WPs, allowing its Chern number to change as the number and chirality of the WPs evolve[20]. This feature provides an opportunity to control the distribution of topological charge between band degeneracies of different dimensionalities. To explore this possibility, we investigate the evolution of the electronic structure of THC-$B_{30}$ under hydrostatic pressure and uniaxial strain along the $a$ and $c$ axes, distinguishing perturbations that preserve or break the sixfold rotational symmetry.

Hydrostatic pressure preserves the $C_6$ rotational symmetry and the symmetry-protected NS. As shown in Fig. 5(a), the original configuration consisting of two $C = -1$ WPs and one $C = +2$ NS remains unchanged over the investigated pressure range from −3 to +1 GPa. Beyond this range, additional band crossings emerge, leading to distinct topological configurations. At −6 GPa, two types of WPs carrying $C = +1$ and $C = -2$ appear along Γ–A, together with their time-reversal-related counterparts. Their combined monopole charge remains compensated by the $C = +2$ NS. Under +3 GPa, the original $C = -1$ WPs persist along Γ–A, while two additional types of WPs with $C = -1$ and $C = +1$ emerge along L–M. The complete BZ contains two original WPs and six symmetry-related WPs of each newly generated type, whose additional charges cancel. Consequently, the NS retains $C = +2$ despite the increased number of WPs (Supplementary Fig. S2). These results demonstrate that hydrostatic pressure can modify the number and charge distribution of WPs while preserving the global charge-neutrality condition.

Uniaxial strain along the *c* axis likewise preserves the $C_6$ symmetry but produces a different evolution of the band crossings [Fig. 5(b)]. Under 3% tensile strain, the original mixed-dimensional configuration is retained (Supplementary Fig. S3). When the tensile strain reaches 5%, a WP with $C = -2$ and two conventional WP with C = +1 emerge along Γ–A, each accompanied by its time-reversal-related counterpart. The total topological charge of all WPs is zero, consistent with the topological charge neutrality theorem, while the NS becomes topologically trivial, with a Chern number of zero. The corresponding surface spectrum exhibits Fermi-arc states connecting the surface projections of these oppositely charged WPs, reflecting the modified bulk topology (Supplementary Fig. S4). In contrast, 1% compressive strain induces a new WP with $C = +1$ along the L–M path [Fig. 5(b)]. Together with the two pre-existing WPs with $C = -1$ along the Γ–A axis, these WPs yield a net topological charge of +4, accounting for their time-reversal-related counterparts. Since no other WPs occur between the two bands throughout the BZ, the NS must carry a $C = -4$ to satisfy the global topological charge neutrality condition. This nontrivial charge compensation is further corroborated by the surface Fermi arcs [Supplementary Fig. S5]. These strain-induced changes illustrate how the topological charge of an extended nodal manifold can respond to the creation and rearrangement of isolated chiral nodes.

A qualitatively different transition occurs under uniaxial strain along the $a$ axis, which breaks the $C_6$ rotational symmetry and lowers the space group from $P6_3$ (No. 173) to $P2_1$ (No. 4). Although the rotational symmetry is reduced, the symmetry-protected NS at the BZ boundary remains intact. Under 1% compressive strain, a single pair of $C = -1$ WPs continues to coexist with a $C = +2$ NS, preserving the original mixed-dimensional charge-compensation mechanism. By contrast, 1% tensile strain generates four unit-charge WPs, comprising two WPs of each chirality. Their monopole charges cancel among themselves, leaving the NS topologically uncharged, with $C_{\mathrm{NS}} = 0$. The corresponding surface-state connectivity changes accordingly, with arc-like states connecting oppositely charged WPs rather than extending towards the NS projection (Supplementary Fig. S6). This transition demonstrates that the symmetry protection of a NS and its nonzero Chern charge are distinct properties: the NS may remain symmetry enforced even when its net topological charge vanishes.

The resulting topological evolution is summarized in Fig. 6. Perturbations that preserve the $C_6$ symmetry can retain the original configuration over a finite range or generate additional conventional and high-charge WPs, with the NS providing the compensating charge required by global neutrality. In contrast, symmetry reduction under *a*-axis strain enables the charge-compensation mechanism to evolve from a mixed-dimensional WP–NS configuration to one governed entirely by WPs of opposite chirality. Therefore, mechanical perturbations provide a route to modifying not only the positions and charges of individual WPs but also the collective topological-charge distribution between zero- and 2D band degeneracies. Furthermore, we confirm that THC-$B_{30}$ retains its dynamical stability under all the aforementioned strain conditions (see Supplementary Fig. S7). Thus, THC-$B_{30}$ offers a promising candidate electronic platform for exploring strain-controlled transitions between distinct charge-neutral topological configurations.

**Discussion**

The coexistence of same-chirality WPs and topologically charged NSs provides an alternative realization of global topological-charge neutrality beyond conventional Weyl-node pairing. Previous theoretical and experimental studies have established this compensation mechanism in acoustic and phononic systems[15-17]. In particular, Ding *et al.* identified six nonmagnetic spinless space groups supporting a single pair of same-chirality unit-charge WPs compensated by a charged NS, with the WPs located at the K and K′ HSPs[15]. Our work extends this classification by systematically examining all 1651 MSGs in both single-valued and double-valued representation frameworks. The resulting 13 compatible type-II MSGs accommodate same-chirality WP pairs not only at HSPs but also along HSLs. This extension broadens the symmetry conditions available for realizing the mixed-dimensional topological configurations and establishes a more systematic basis for their identification in crystalline materials.

Beyond symmetry classification, identifying the first predicted electronic realization requires a material in which the relevant band crossings remain sufficiently isolated to distinguish the minimal unpaired topological configuration. The predicted chiral boron allotrope THC-$B_{30}$ provides such a candidate in the spinless electronic description. Its weak atomic SOC favours the preservation of the symmetry-protected crossings, while its electronic structure contains a single pair of $C = -1$ WPs along the Γ–A direction and a $C = +2$ NS on the $k_z = \pi$ plane between the relevant adjacent bands. The corresponding surface-state arcs connect the projections of the WPs to that of the NS, extending the previously demonstrated mixed-dimensional bulk–boundary correspondence in artificial acoustic systems to a predicted electronic crystalline platform[16, 17]. Moreover, the inversion relationship between the two structural enantiomers reverses the associated chiral charges, establishing a direct connection between structural handedness and momentum-space topology.

An additional implication of this work is that the topological charge of a symmetry-protected NS need not remain fixed as the surrounding band structure evolves. Our strain calculations show that conventional and higher-charge WPs can emerge or rearrange while preserving global charge neutrality, accompanied by changes in the compensating charge of the NS. More importantly, breaking the sixfold rotational symmetry can transform the original WP–NS charge-compensation configuration into one in which the charges are balanced entirely among oppositely charged WPs, even though the nodal plane itself persists. This distinction between symmetry-enforced degeneracy and topological charge provides a useful perspective on the evolution of mixed-dimensional gapless phases. The symmetry framework developed here may guide the identification of related configurations in other weak-SOC electronic materials and, more broadly, in phononic and photonic systems. Together, these findings establish a route towards realizing and manipulating same-chirality Weyl pairs through the collective topology of nodal objects with different dimensionalities.

In summary, we have established a symmetry framework for realizing a minimal unpaired mixed-dimensional topological composite comprising a single charged NS and two same-chirality unit-charge WPs. By systematically examining all 1651 MSGs with and without SOC, we identify 13 compatible type-II MSGs and predict the chiral boron allotrope THC-$B_{30}$ as a candidate electronic realization. Its two $C = -1$ WPs and single $C = +2$ NS satisfy global topological-charge neutrality and exhibit unconventional Fermi arcs connecting their surface projections. Furthermore, strain-induced topological phase transitions redistribute the chiral charges among WPs and the NS, enabling distinct charge-compensation configurations. Our findings provide a symmetry-guided route to realizing and controlling same-chirality Weyl pairs and extend the exploration of topological semimetals towards charge-neutral composites of momentum-space objects with different dimensionalities.

## Methods

**Computational methods:** The structural and electronic properties of THC-$B_{30}$ were investigated through density functional theory (DFT) calculations using the Vienna *Ab initio* Simulation Package (VASP)[41]. The electron–ion interactions were described within the projector augmented-wave (PAW) formalism[42], while exchange–correlation effects were treated using the Perdew–Burke–Ernzerhof (PBE) functional within the generalized gradient approximation (GGA)[43]. A plane-wave cutoff of 520 eV and a $6 \times 6 \times 8$ Monkhorst–Pack *k*-point meshes[44] were adopted, with the total-energy and atomic-force convergence thresholds set to $10^{-6}$ eV and $10^{-3}$ eV/Å, respectively. Phonon spectra were obtained using the finite-displacement approach implemented in PHONOPY package[45] with 2 × 2 × 2 supercells. Finite-temperature structural stability was examined by *ab initio* molecular dynamics (AIMD) simulations in the canonical (NVT) ensemble at 800 K for 8 ps, employing a Nosé–Hoover thermostat[46]. The elastic stiffness tensor $C_{ij}$ were evaluated via using energy-stress method[47]. The tight-binding Hamiltonian was constructed from maximally localized Wannier functions using Wannier90[48] and employed to calculate the Berry curvature, Weyl-point Chern numbers, and surface electronic structures using WannierTools Packages[49].

## Data availability

The data supporting the findings are displayed in the main text and the Supplementary Information. All raw data are available from the corresponding authors upon request.

## Acknowledgements:

This work was supported by the National Natural Science Foundation of China (Grants No. 12304202), the Hebei Natural Science Foundation (Grant No. A2023203007), the Science Research Project of Hebei Education Department (Grant No. BJK2024085), and the Cultivation Project for Basic Research and Innovation of Yanshan University (No. 2022LGZD001).

## Author contributions:

X.-J. Gao proposed the THC-$B_{30}$ structure and performed the corresponding DFT calculations. K.-X. Pang carried out the symmetry analysis. Y. Gao conceived the original ideas. X.-J. Gao, Y.-Y. Bai and Y. Ge analyzed the results. Y. Gao writing-reviewing, conceptualization, supervision, and project administration. All authors discussed the results and commented on the manuscript at all stages.

## Competing interests

The authors declare no competing financial interests.

# Figures and Tables

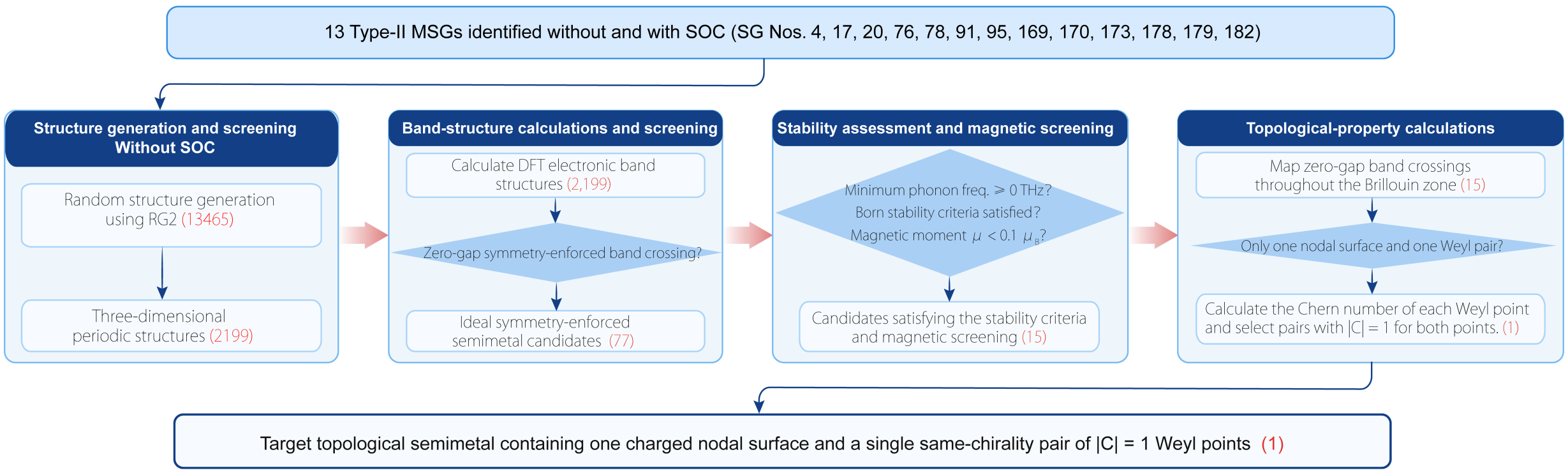

**Figure 1 | Symmetry-guided high-throughput discovery of mixed-dimensional topological semimetals.** Schematic workflow for identifying boron-based materials hosting a single charged nodal surface and a pair of same-chirality unit-charge Weyl points within the 13-candidate type-II magnetic space groups (SGs). The successive screening stages yield 2,199 three-dimensional structures from 13,465 initial candidates, followed by 77 symmetry-compatible semimetals, 15 structurally stable nonmagnetic candidates, and finally one target material, THC-$B_{30}$. Red numbers denote the number of structures retained at each screening stage.

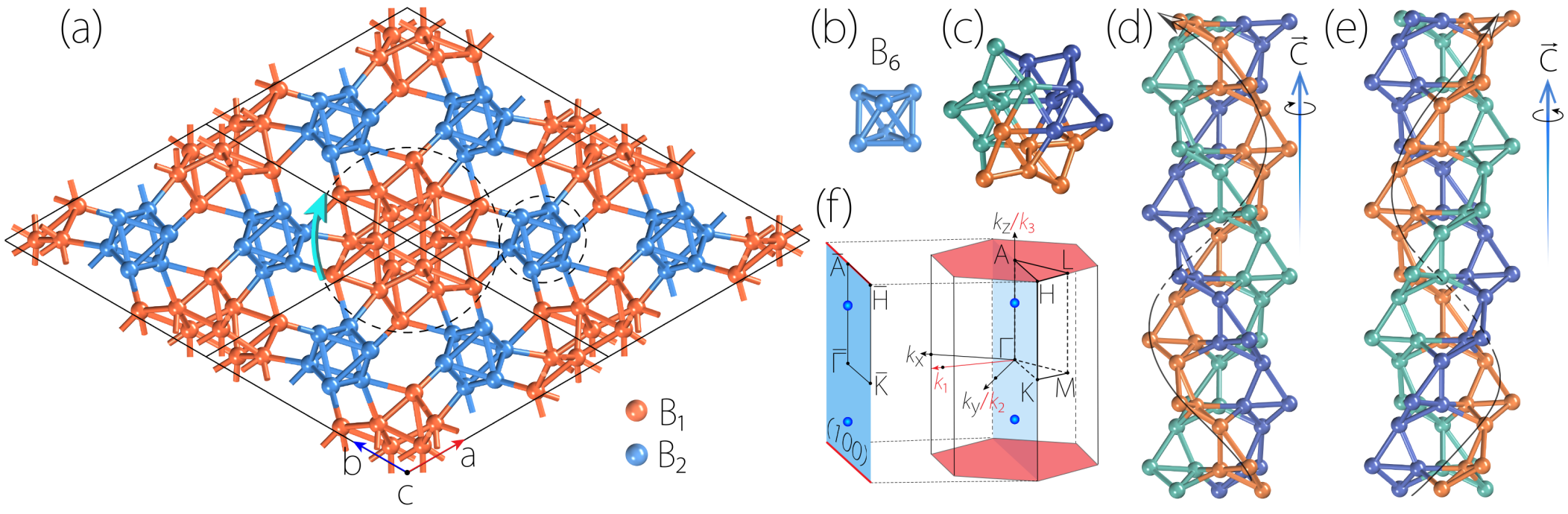

**Figure 2 | Crystal structure and Brillouin zone of chiral THC-$B_{30}$.** (a) Top view of the THC-$B_{30}$ crystal framework, composed of interconnected bilayer triangular $B_6$ units and triangular helical boron chains. (b) Bilayer triangular $B_6$ unit formed by two stacked and interconnected boron triangles. (c) Triangular helical-chain unit comprising three helically intertwined boron chains. (d,e) Side views of the left- and right-handed helical units, respectively. (f) Bulk Brillouin zone and (100) surface projection. Red planes represent the nodal surface at $k_z = \pi$, while blue dots indicate the Weyl points.

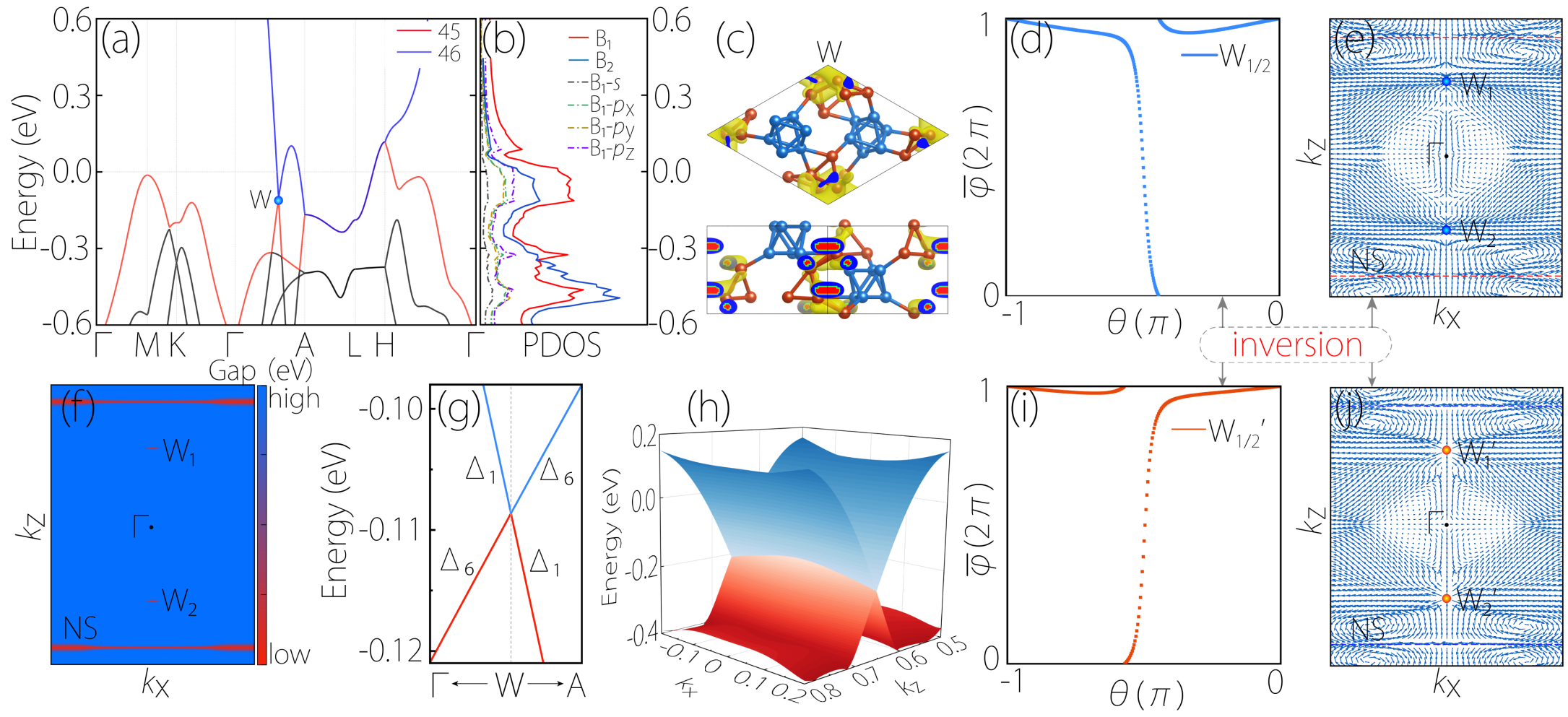

**Figure 3 | Electronic structure and mixed-dimensional topological charge compensation in THC-B₃₀.** (a) Band structure near the Fermi level, highlighting the 45th and 46th bands. (b) Orbital-projected density of states. (c) Charge-density distribution at the W point. (d) Wannier charge-centre (WCC) evolution for $W_1$ and $W_2$, both with $C = -1$. (e) Berry-curvature distribution of the left-handed structure. (f) Projected band-gap map, with red dots and the red line denoting the Weyl points and nodal surface, respectively. (g) Band crossing along Γ–A, labelled by the irreducible representations $\Delta_1$ and $\Delta_6$. (h) Three-dimensional band dispersion showing the nodal surface at $k_z = \pi$. (i,j) WCC evolution and Berry-curvature distribution of the right-handed structure, with $C = +1$ for both Weyl points.

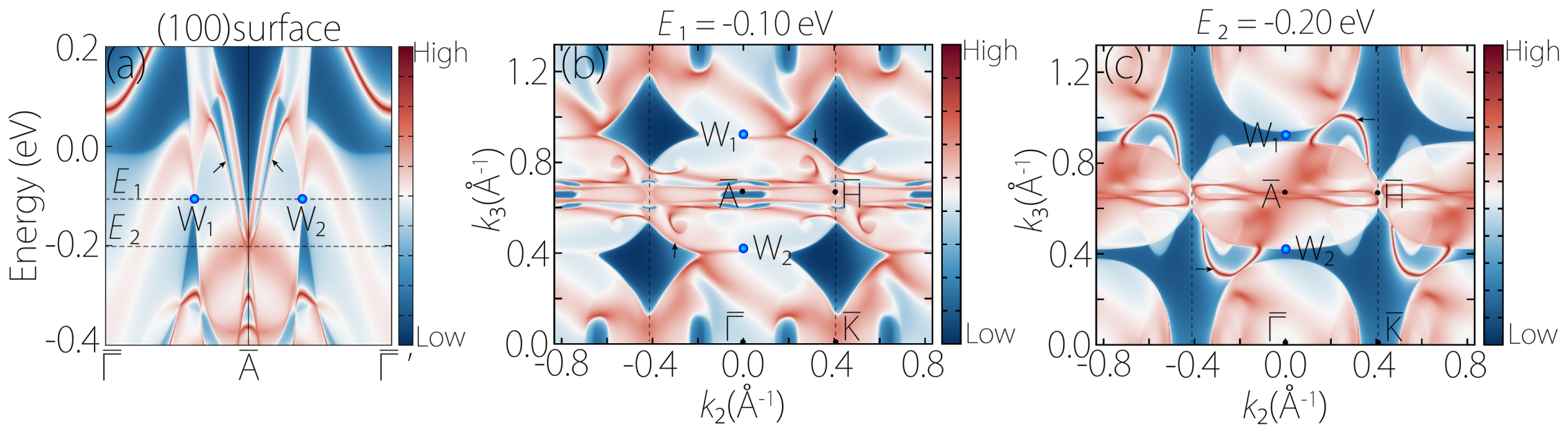

**Figure 4 | Unconventional Fermi arcs connecting Weyl points and a charged nodal surface.** (a) Surface spectral function along $\bar{\Gamma}-\bar{\mathrm{A}}-\bar{\Gamma}'$ on the (100) surface. (b,c) Constant-energy surface spectra at $E_1=-0.10$ and $E_2=-0.20$ eV, respectively. Blue dots denote the projected $C=-1$ Weyl points $W_1$ and $W_2$. The nodal surface projects along $\bar{\mathrm{A}}-\bar{\mathrm{H}}$; arrows indicate the connecting surface-state arcs.

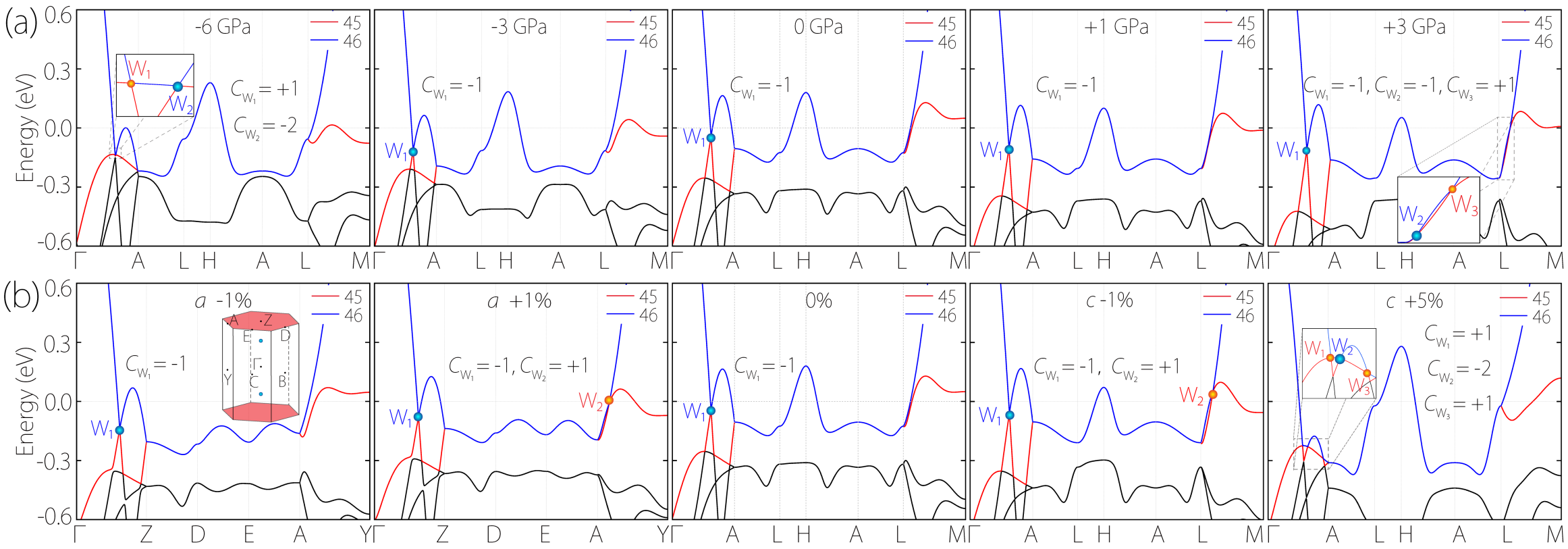

**Figure 5 | Evolution of Weyl band crossings under hydrostatic pressure and uniaxial strain.** (a) Band structures under hydrostatic pressures of −6, −3, 0, +1, and +3 GPa (left to right). Additional Weyl crossings appear at −6 and +3 GPa, while the original configuration persists at intermediate pressures. (b) Band structures under *a*-axis strains of −1% and +1%, zero strain, and *c*-axis tensile strains of +3% and +5% (left to right). The inset shows the Brillouin zone after *a*-axis symmetry reduction. Blue and orange dots denote negatively and positively charged Weyl points, respectively; the indicated Chern numbers label the crossings.

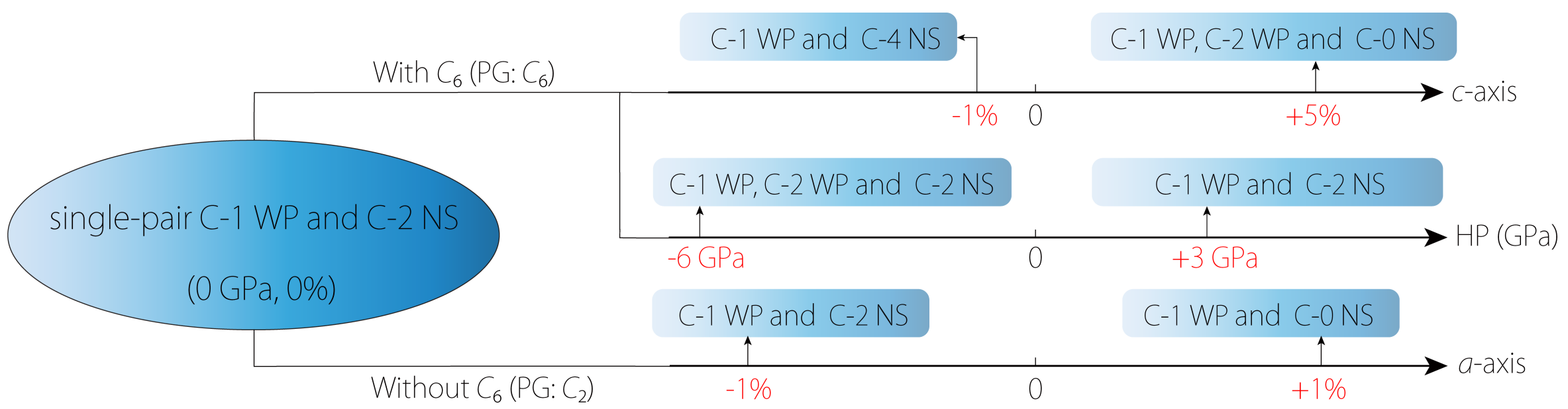

**Figure 6 | Topological phase diagram of THC-$B_{30}$ under strain.** Schematic charge configurations under *c*-axis strain (top), hydrostatic pressure (middle), and *a*-axis strain (bottom). The unstrained state contains two $C = -1$ Weyl points and one $C = +2$ nodal surface. Labels indicate the Weyl-point and nodal-surface charges in the strained phases. $C_6$ symmetry is preserved under *c*-axis strain and hydrostatic pressure and broken under *a*-axis strain.

**Table 1 | Magnetic space groups compatible with a single charged nodal surface (NS) and one pair of same-chirality unit-charge Weyl points (WPs).** Classification of the 13 type-II magnetic space groups (SGs) identified as compatible with the targeted mixed-dimensional topological configuration, considering both the absence and presence of spin–orbit coupling (SOC). The tables specify the candidate SGs, allowed high-symmetry points (HSPs) and high-symmetry lines (HSLs), relevant point-group (PG) symmetries, and irreducible representations or corepresentations (IRRs) associated with the nodal-surface and Weyl-point crossings. The corresponding absolute Chern-charge magnitudes are $|C|_{\mathrm{NS}} = 2$ and $|C|_{\mathrm{W}} = 1$. Detailed symmetry generators and representation assignments are provided in Supplementary Table S1.

| SG Nos. | PG | NS: HSPs / IRRs | WPs: HSPs or HSLs / IRRs |
|---|---|---|---|
| **Without SOC (single-valued representations)** | | | |
| **4** | $C_2$ | Z, C, D, E: $\{R_2, R_4\}$ | ΓZ, BD, YC, AE: $\{R_1\},\{R_2\}$ |
| **17, 20** | $D_2$ | **17:** Z, U, T, R: $\{R_5\}$; **20:** Z, T: $\{R_5\}$ | **17:** ΓY, XS, ΓX, YS, ΓZ, YT, SR, XU: $\{R_1\},\{R_2\}$; **20:** ΓZ, YT, ΓY, ΓΔ, YF, YC: $\{R_1\},\{R_2\}$ |
| **76, 78** | $C_4$ | Z, A: $\{R_2, R_8\}$ or $\{R_4, R_6\}$ | ΓZ, MA: $\{R_1\},\{R_2\}$; $\{R_1\},\{R_4\}$; $\{R_2\},\{R_3\}$; $\{R_3\},\{R_4\}$ |
| **91, 95** | $D_4$ | Z, A $R_6$ or $R_7$ | ΓZ, MA $\{R_1\},\{R_2\}$; $\{R_1\},\{R_4\}$; $\{R_2\},\{R_3\}$; $\{R_3\},\{R_4\}$ |
| **169, 170, 173** | $C_6$ | A: $\{R_2, R_{12}\}$; $\{R_4, R_{10}\}$; $\{R_6, R_8\}$ | K, ΓA: $\{R_2, R_3\}$; $\{R_1\},\{R_2\}$; $\{R_1\},\{R_6\}$; $\{R_2\},\{R_3\}$; $\{R_3\},\{R_4\}$; $\{R_4\},\{R_5\}$; $\{R_5\},\{R_6\}$ |
| **178, 179, 182** | $D_6$ | A: $R_7$; $R_8$; $R_9$ | K, ΓA: $R_3$; $\{R_1\},\{R_2\}$; $\{R_1\},\{R_6\}$; $\{R_2\},\{R_3\}$; $\{R_3\},\{R_4\}$; $\{R_4\},\{R_5\}$; $\{R_5\},\{R_6\}$ |
| **With SOC (double-valued representations)** | | | |
| **4** | $C_2$ | Z, C, D, E: $\{R_5, R_5\}$ or $\{R_7, R_7\}$ | Γ, B, Y, A, ΓZ, BD, YC, AE: $\{R_2, R_4\}$ or $\{R_2\},\{R_4\}$ |
| **17, 20** | $D_2$ | **17:** Z, U, T, R: $\{R_5, R_6\}$ or $\{R_7, R_8\}$; **20:** Z, T $\{R_5, R_6\}$ or $\{R_7, R_8\}$ | **17:** Γ, Y, X, S, ΓY, XS, ΓX, YS, ΓZ, YT, SR, XU: $\{R_2, R_4\}$ or $\{R_2\},\{R_4\}$; **20:** Γ, Y, S, ΓZ, YT, ΓY, ΓΔ, YF, YC: $\{R_2, R_4\}$ or $\{R_2\},\{R_4\}$ |
| **76, 78** | $C_4$ | Z, A: $\{R_1, R_1\}$; $\{R_3, R_7\}$; $\{R_5, R_5\}$ | Γ, M, ΓZ, MA: $\{R_2, R_8\}$; $\{R_4, R_6\}$; $\{R_2\},\{R_4\}$; $\{R_2\},\{R_8\}$; $\{R_4\},\{R_6\}$; $\{R_6\},\{R_8\}$ |
| **91, 95** | $D_4$ | Z, A: $\{R_8, R_{10}\}$; $\{R_9, R_{11}\}$; $R_{14}$ | Γ, M, ΓZ, MA: $R_6$; $R_7$; $\{R_2\},\{R_4\}$; $\{R_2\},\{R_8\}$; $\{R_4\},\{R_6\}$; $\{R_6\},\{R_8\}$ |
| **169, 170, 173** | $C_6$ | A: $\{R_1, R_1\}$; $\{R_3, R_{11}\}$; $\{R_5, R_9\}$; $\{R_7, R_7\}$ | K, ΓA: $\{R_2, R_6\}$; $\{R_2\},\{R_{12}\}$; $\{R_2\},\{R_4\}$; $\{R_4\},\{R_6\}$; $\{R_6\},\{R_8\}$; $\{R_8\},\{R_{10}\}$; $\{R_{10}\},\{R_{12}\}$ |
| **178, 179, 182** | $D_6$ | A: $\{R_{10}, R_{11}\}$; $\{R_{12}, R_{13}\}$; $R_{14}$; $R_{15}$ | K, ΓA: $R_6$; $\{R_2\},\{R_{12}\}$; $\{R_2\},\{R_4\}$; $\{R_4\},\{R_6\}$; $\{R_6\},\{R_8\}$; $\{R_8\},\{R_{10}\}$; $\{R_{10}\},\{R_{12}\}$ |